# Label-free colorectal cancer screening using deep learning and spatial light interference microscopy (SLIM)

*Jingfang "Kelly" Zhang[1,3], Yuchen R. He[1,2,3], Nahil Sobh[1,3], and Gabriel Popescu[1,2,3,4]*

[1] Quantitative Light Imaging Laboratory

[2] Department of Electrical and Computer Engineering

[3] Beckman Institute of Advanced Science and Technology

[4] Department of Bioengineering

University of Illinois at Urbana-Champaign, 405 N. Matthews Avenue, Urbana, IL 61801, USA.

sobh@illinois.edu

gpopescu@illinois.edu

**One Sentence Summary:** Combining SLIM and deep learning, we classify normal and cancer glands with >97% accuracy, which presents an opportunity for screening applications.

**Abstract:**

Current pathology workflow involves staining of thin tissue slices, which otherwise would be transparent, followed by manual investigation under the microscope by a trained pathologist. While the hematoxylin and eosin (H&E) stain is well-established and a cost-effective method for visualizing histology slides, its color variability across preparations and subjectivity across clinicians remain unaddressed challenges. To mitigate these challenges, recently we have demonstrated that spatial light interference microscopy (SLIM) can provide a path to intrinsic, objective markers, that are independent of preparation and human bias. Additionally, the sensitivity of SLIM to collagen fibers yields information relevant to patient outcome, which is not available in H&E. Here, we show that deep learning and SLIM can form a powerful combination for screening applications: training on 1,660 SLIM images of colon glands and validating on 144 glands, we obtained a benign vs. cancer classification accuracy of 99%. We envision that the SLIM whole slide scanner presented here paired with artificial intelligence algorithms may prove valuable as a pre-screening method, economizing the clinician's time and effort.

**Introduction**

Quantitative phase imaging (QPI) [1] has emerged as a powerful label-free method for biomedical applications [2] More recently, due to its high sensitivity to tissue nanoarchitecture and quantitative output, QPI has been proven valuable in pathology [3, 4]. Combining spatial light interference microscopy (SLIM, [5, 6]) and dedicated software for whole slide imaging (WSI) allowed us to demonstrate the value of tissue refractive index as an intrinsic marker for diagnosis and prognosis [7-14]. So far, we have used various metrics derived from the QPI map to obtain clinically relevant information. For example, we found that translating the data into tissue scattering coefficients can be used to predict disease recurrence after prostatectomy. SLIM's sensitivity to collagen fibers proved useful in the diagnosis and prognosis of breast cancer [12, 14, 15]. While this approach of "feature engineering" has the advantage of providing physical significance to the computed markers, it only covers a limited range of parameters available from our data. In other words, it is likely that certain useful parameters are never computed at all. This restricted analysis is likely to limit the ultimate performance of our procedure.

Recently, artificial intelligence (AI) has received significant scientific interest from the biomedical community [16-20]. In image processing, AI provides an exciting opportunity for boosting the amount of information from a given set of data, with high throughput [20]. In contrast to feature engineering, a deep convolution neural network computes an exhaustive number of features associated with an image, which is bound to improve the performance of the task at hand. Here, we apply, for the first time to our knowledge, SLIM and AI to classify colorectal tissue into cancer and benign.

Genetic mutations over the course of 5-10 years leads to the development of colorectal cancer from benign adenomatous polyps [21]. Early diagnosis promotes disease-specific mortality.

Thus, early diagnosed cancers (still localized) have a 89.8% 5-year survival rate compared to a 12.9% 5-year survival rate for patients with distant metastasis or late stage-disease [22]. Colonoscopy is the preferred form of screening in the U.S. From 2002 to 2010, the percentage of persons in the age group of 50-75 years who underwent colorectal cancer screening increased from 54% to 65% [23]. Out of all individuals undergoing colonoscopy, the prevalence of adenoma is 25 - 27%, and the prevalence of high-grade dysplasia and colorectal cancer is 1 - 3.3% [24, 25]. As current screening methods cannot distinguish adenoma from a benign polyp with high accuracy, a biopsy or polyp removal is performed in 50% of all colonoscopies [26]. A pathologist examines the excised polyps to determine if the tissue is benign, dysplastic, or cancerous.

New technologies for quantitative and automated tissue investigation are necessary to reduce the dependence on manual examination and provide large-scale screening strategies. As a successful precedent, the Papanicolou test (pap smear) for cervical cancer screening has been augmented by the benefits of computational screening tools [27]. The staining procedure, which is critical to the proper operation of such systems, is designed to match calibration thresholds [28].

We used a SLIM-based tissue scanner in combination with AI to classify cancer and benign cases. We demonstrate the clinical value of the new method by performing automatic colon screening, using *intrinsic* tissue markers. Importantly, such a measurement does not require staining or calibration. Therefore, in contrast to current staining markers, signatures developed from the phase information can be shared across laboratories and instruments, without modification.

**Results and methods**

*SLIM whole slide scanner*

Our label-free SLIM scanner, consisting of dedicated hardware and software, is described in more detail in [29]. Figure 1 illustrates the SLIM module (Cell Vista SLIM Pro, Phi Optics, Inc.), which outfits an existing phase contrast microscope. In essence, SLIM works by making the ring in the phase contrast objective pupil *tunable*. In order to achieve this, the image outputted by a phase contrast microscope is Fourier transformed at a plane of a spatial light modulator (SLM), which produces pure phase modulation. At this plane, the image of the phase contrast ring is perfectly matched to the SLM phase mask, which is shifted in increments of 90º (Fig. 1). From the four intensity images that correspond to the ring phase shifts, the quantitative phase image is retrieved uniquely at each point in the field of view. Figure 2 shows examples of SLIM images associated with tissue cores and glands for cancer and normal colon cases.

The SLIM tissue scanner can acquire the four intensity images, process them, and display the phase image, all in real-time. This is possible due to the novel acquisition software that seamlessly combines CPU and GPU processing [29]. The SLIM phase retrieval computation occurs on a separate thread while the microscope stage moves to the next position. Scanning large fields of view, e.g., entire microscope slides, and assembling the resulting images into single files required the development of new dedicated software tools [29]. The final SLIM images (Fig. 2) are displayed in real-time at up to 15 frames per second, as limited by the spatial light modulator refresh rate, which is 4X faster.

*Deep learning model*

Our deep learning model leverages pretrained *convnet*s for which we used a total of 1884 gland images. The data are split into three sets, with 1660 images used as a training set , while 144 gland images used as validation dataset, and the remaining 40 gland images used as the "hidden" test dataset, which reports the final performance of our deep learning classifier. We used a transfer learning approach to build our deep learning classifier. This approach is especially useful when there is a limited amount of data to train a model. We selected the VGG16 deep network trained on a large dataset (ImageNet over 1.6 M images of various sizes and 1000 classes – See Figure 3). Among the long list of pretrained models like ResNet, Inception, Inception-ResNet, Xception, MobileNet and others, we chose the VGG16 network, due to its rich features extraction capabilities (see Fig. 3 network and Ref. [30]). The 138M parameters of VGG16 uses over 528 MB in storage and has only 16 layers. We reuse the VGG16's parameters in the convolutional, the first five blocks of the network, to extract the rich features that are hidden within each gland image. The "top layer" of the VGG16 network is replaced by several fully connected layers and a final sigmoid nonlinear activation unit. The output of the nonlinear activation function is class predicted by the network. The predicted classification is fed into a "cross-entropy" loss function. Stochastic gradient descent methods are used to update the weights of the new network in two steps, as follows. The training of our network is carried out in two steps. In the first step, we import the VGG16 weights and replace the top layer (consisting of three fully connected layers and a softmax classifier) with our fully connected layers and a sigmoid for binary classification. During this first step we "freeze" the weights of convolutional layers of the VGG16 network and update the weights of newly added top layers. In the second step and final step, we "unfreeze" the weights in the convolutional layers and fine tune all the weights (convolutional and top layers). The original input image size of the VGG16 was 224x224x3, we increased it to 256x256x3. Since our images are in gray scale and that the VGG16 only accepts RGB image as input, we had to copy each image three times and place the threes copies in R, G, and B channels. The new fully connected (FC) layers have 2048 units each, followed by a dense layer with 256 units, a dropout of 0.5 is used immediately after the first

2048 FC layer. This selection of FCs and dropout resulted in best performance on our validation and test sets.

*Model accuracy and loss*

Model accuracy and losses are shown in Fig. 4. First, note that the shape of loss curves is a good proxy for assessing the "underfitting", "overfitting", and "just-right" models. In general, a deep learning model is classified as "underfitting", when it is not efficient use of all training dataset. In this case, the training loss curve exhibits a non-zero constant loss value beyond a certain epoch value. In a similar fashion, a deep learning model is said to "overfit" the training data, when the training loss curve keeps decreasing, while the validation loss metric stalls and then starts to increase. Both "underfitting" and "overfitting" are signs of non-generalizability. On the other hand, the 'just-right" deep learning models, the training and validation loss functions tend to follow each other closely and converge towards zero or very small values. We stopped the training at epoch 60 (this is known as training by early stopping criteria ), where the network is no longer able to generalize (i.e., the validation loss started to increase after a specific epoch value). Early stopping criteria are implemented by saving the trained weights of our network where validation loss is lowest during all the training cycle. During our training exercise, the "best" model is saved when the validation accuracy highest is 0.98 at epoch 36.

*ROC, AUC, and classification reports for validation and test*

Receiver Operating Characteristics (ROC) curve and Area Under the Curve (AUC) scores are two metrics used in reporting the performance of our network on both the validation and the test datasets (Fig. 5). The ROC curve displays the performance of our deep learning classifier at various thresholds. The two parameters plotted on the ROC axes are the true positive rate, along the y-axis, and false positive rate along the x-axis. Figure 5a shows the ROC curve for the validation and test. The AUC measure on the validation set is 0.98 and 0.99 in the test set, also show in Fig. 5d. The accuracy for both the validation and test dataset was 97%.

*Confusion matrix for validation and test*

The confusion matrix provides a quantitative measure performance of binary classifier. There are two classes in our confusion matrix: "normal" or "cancer" gland. A confusion matrix has two types of errors: Type I error is False Positive, where "normal" is classified as "cancer". Type II error is False Negative, where "cancer" is classified as "normal". For a perfect classifier, its confusion matrix is diagonal, which means it will only have True Negatives and True Positives.

The confusion matrices for validation and test datasets are shown in Figures 6a and 6b, respectively. In the first row of Fig. 6a, 69 out of the 72 "normal" instances are correctly predicted or True Negatives; and 3 out of the 72 are wrongly predicted as "cancer" or False Positives (Type I error). In the second row of Fig. 6a, 71 out of the 72 "cancer" instances are correctly predicted as "cancer" or True Positives; and 1 out of the 72 "cancer" instances is wrongly predicted as "normal" or False Negative (Type II error). In the first row of Fig. 6b, all the 20 "normal" instances are correctly predicted as "normal" and none of them is wrongly predicted as "cancer". In the second row of Fig. 6b, 19 out of the 20 instances are correctly predicted as "cancer" and only 1 out of the 20 instances is wrongly predicted as "normal" or False Negative (Type II error).

**Summary and discussion**

In summary, we showed that applying AI (deep transfer learning) to SLIM images yields excellent performance in classifying cancers and benign tissue. The 98% (validation dataset) and 99% (test dataset) overall glandular scale accuracy, defined as area under the ROC curve suggest that this approach may prove valuable especially for screening applications. The SLIM module can be implemented to existing microscopies already in use in the pathology laboratories around the world. Thus, it is likely that this new tool can be easily adopted at a large scale as a prescreening tool, enabling the pathologist to screen trough cases fast. This approach can be applied to more difficult tasks in the future, such as quantifying the aggressiveness of the disease [12], and can be used for other types of cancer, with proper optimization of the network.

It has been shown in a different context that the inference step can be implemented into the SLIM acquisition software [Kandell et all., under review]. Because the inference is faster than the acquisition time of a SLIM frame and can also be performed in parallel, we anticipate that the classification can be performed in real time. The overall throughput of the SLIM tissue scanner is comparable with that of commercial whole slide scanners that only perform bright field imaging on stained tissue sections [31]. In principle, it is possible to have the result of classification, with areas of interest highlighted for the clinician, all done as soon as the scan is complete, in a couple of minutes. In the next phase of this project, we plan to work with clinicians to further assess the performance of our classifier against experts.

**Acknowledgements**

We are grateful to Mikhail Kandel, Shamira Sridharan, and Andre Balla for imaging, annotating and diagnosing the tissues used in this study.

This work was funded by NSF 0939511, R01 GM129709, R01 CA238191, R43GM133280-01

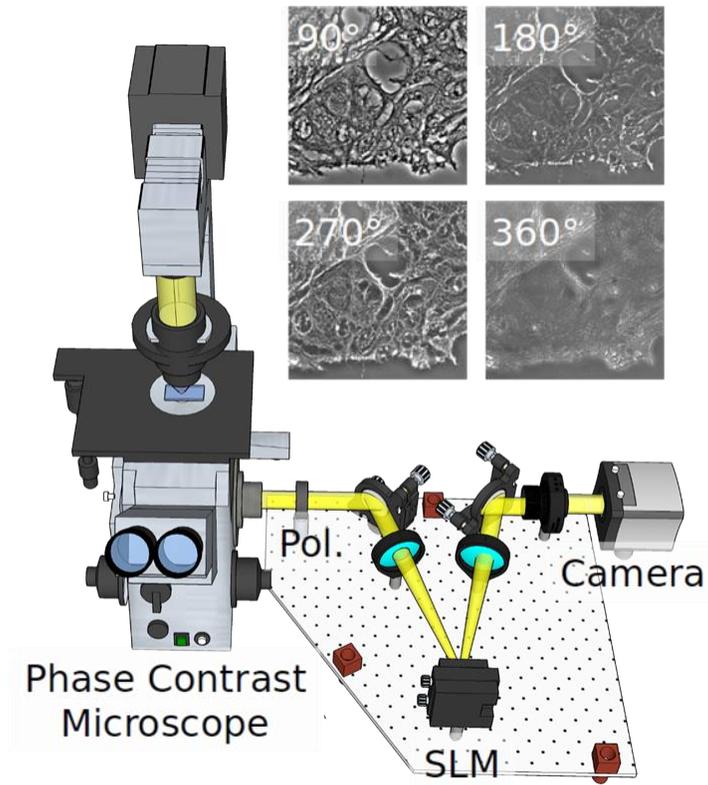

Fig. 1. SLIM system implemented as add-on to an existing phase contrast microscope. Pol-polarizer, SLM-spatial light modulator. The four independent frames corresponding to the 4 phase shifts imparted by the SLM are shown for a tissue sample.

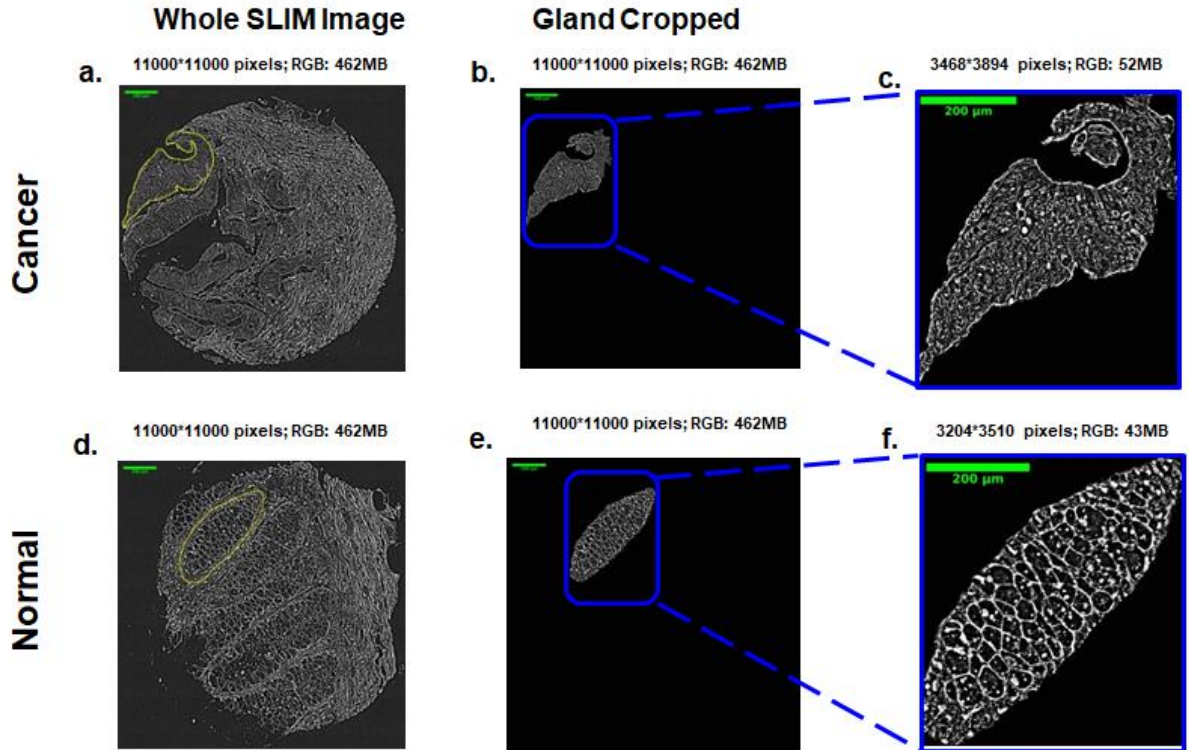

Fig. 2. Examples of cancer (a-c) and normal (d-f) tissue from a microarray of colon patients. The classification is performed at the glandular scale, which are manually annotated for ground truth as illustrated in b-c and e-f.

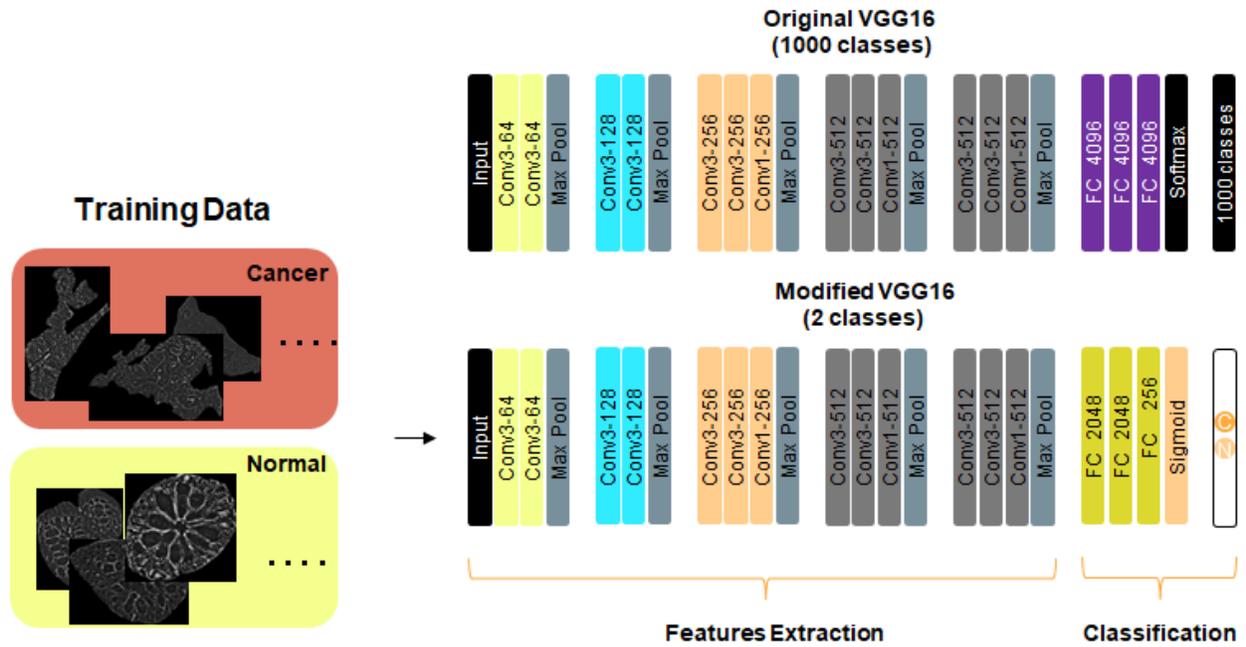

Fig. 3. Modified VGG16 network. Input image size is 256 x 256 x3. A pad of length 1 is added before each Max Pool layer. Conv1 : Convolutional layer with 1x1 filter; Conv3 : Convolutional layer with 3x3 filter; Max Pool: Maximum pooling layer over 2x2 pixels (stride=2); All hidden layers are followed by RELU activation. First FC layer is followed by 0.5 dropout

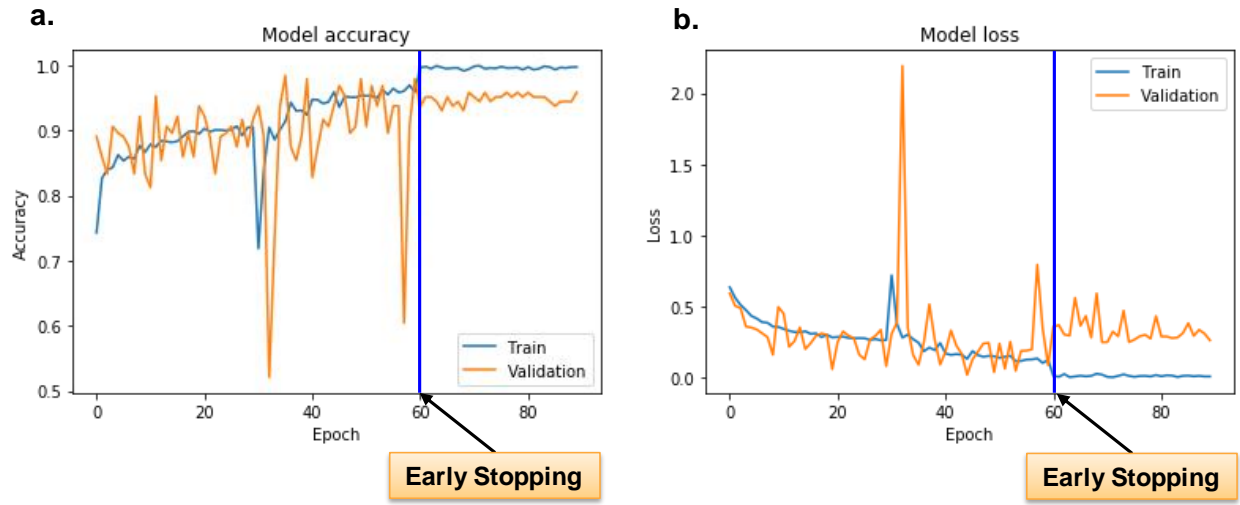

Fig. 4. **Model accuracy and loss.** Training and validation accuracy and loss, with moderate data augmentation. a.) The training accuracy ranges from 0.7186 to 0.9695, while the validation accuracy from 0.5208 to as high as 0.9844. b.) The training loss ranges from 0.1029 to 0.7192, while the validation loss from 0.019 to 2.1918. The biggest validation loss, 2.1918, occurs at the epoch 33, while the validation accuracy hits the lowest level, 0.5208. The blue line in figure a and figure b represents early stopping point since after epoch 60, the validation loss is no longer improving. Only the best model is saved during the whole training.

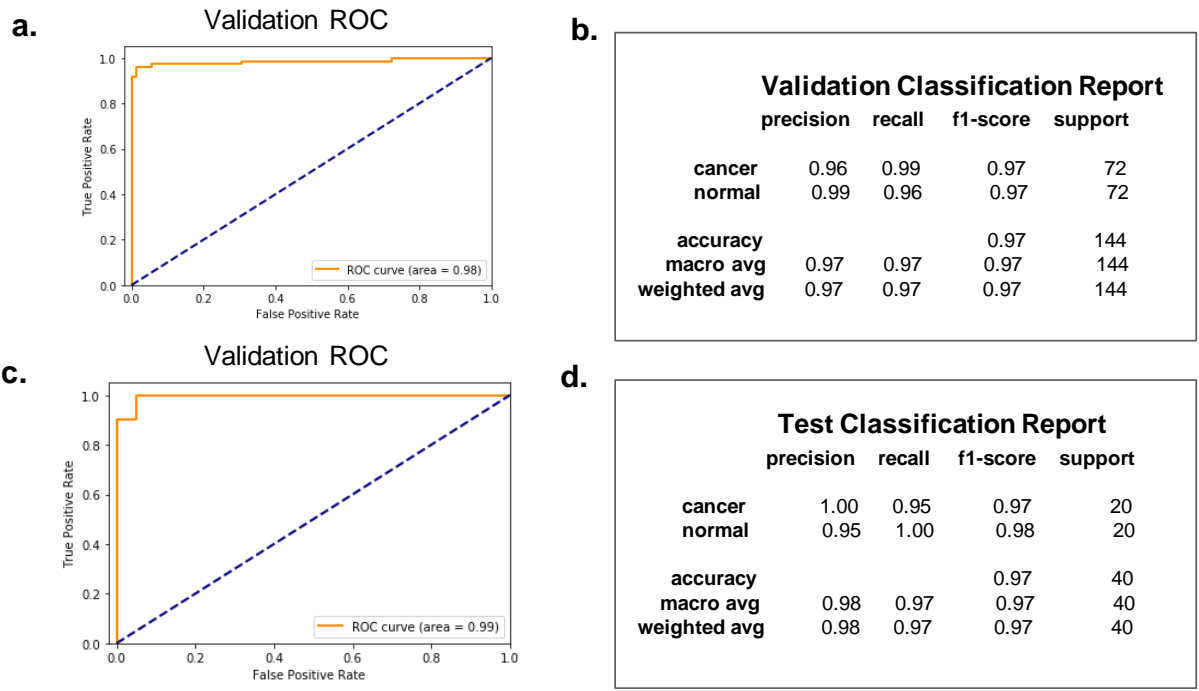

**Fig. 5** ROC (Receiver Operating Characteristics) curve, with AUC (Area Under The Curve), and classification reports respectively for the validation dataset and the test dataset. The AUC score is 0.98 for the validation dataset, and 0.99 for the test dataset, as indicated. The two classes, cancer and normal, have balanced support for both the validation dataset and the test dataset, with the validation dataset providing 72 actual occurrences for each class and the test dataset providing 20 actual occurrences for each class. The accuracy hits 97% for both the validation and the test.

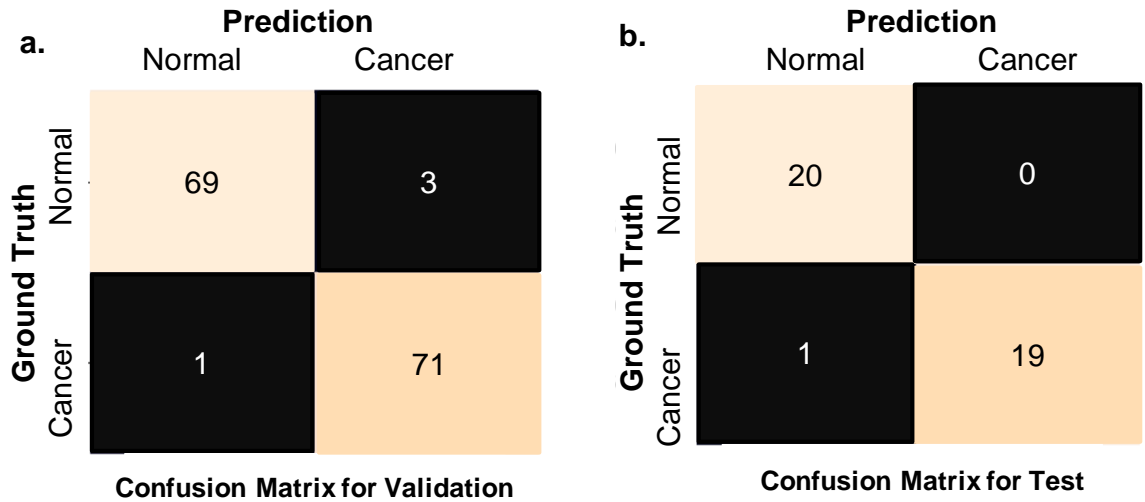

**Fig. 6 Confusion Matrix for Validation (a) and Test (b).** a.) For class cancer, 71 of the 72 images were correctly classified as cancer, while for class normal, 69 of the 72 images were correctly classified as normal. b.) For class cancer, 19 of the 20 images were correctly classified as cancer, while for class normal, all the 20 images are correctly classified as normal.